\documentclass[aps,prl,twocolumn,twoside,amsmath,amssymb,amsfonts,10pt]{revtex4}

\usepackage{graphicx,subfigure}
\usepackage{color}
\usepackage{epstopdf}
\usepackage{CJK}

\begin{document}
%\begin{CJK*}{GBK}{song}

\title{Universal spatiotemporal dynamics of spontaneous superfluidity breakdown in the presence of synthetic gauge fields}

\author{Shuyuan Wu$^{1,2}$}
\author{Xizhou Qin$^{1}$}
\author{Jun Xu$^{1,3}$}
\author{Chaohong Lee$^{1,2,4,}$}\altaffiliation{Email: lichaoh2@mail.sysu.edu.cn; chleecn@gmail.com}

\affiliation{$^{1}$TianQin Research Center \& School of Physics and Astronomy, Sun Yat-Sen University (Zhuhai Campus), Zhuhai 519082, China}
\affiliation{$^{2}$State Key Laboratory of Optoelectronic Materials and Technologies, Sun Yat-Sen University (Guangzhou Campus), Guangzhou 510275, China}
\affiliation{$^{3}$Center of Experimental Teaching for Common Basic Courses, South China Agriculture University, Guangzhou 510642, China}
\affiliation{$^{4}$Synergetic Innovation Center for Quantum Effects and Applications, Hunan Normal University, Changsha 410081, China}

\date{\today}

\begin{abstract}
  According to the famous Kibble-Zurek mechanism (KZM), the universality of spontaneous defect generation in continuous phase transitions (CPTs) can be understood by the critical slowing down.
  In most CPTs of atomic Bose-Einstein condensates (BECs), the universality of spontaneous defect generations has been explained by the divergent relaxation time associated with the nontrivial gapless Bogoliubov excitations.
  However, for atomic BECs in synthetic gauge fields, their spontaneous superfluidity breakdown is resulted from the divergent correlation length associated with the zero Landau critical velocity.
  Here, by considering an atomic BEC ladder subjected to a synthetic magnetic field, we reveal that the spontaneous superfluidity breakdown obeys the KZM.
  The Kibble-Zurek scalings are derived from the Landau critical velocity which determines the correlation length.
  In further, the critical exponents are numerically extracted from the critical spatial-temporal dynamics of the bifurcation delay and the spontaneous vortex generation.
  Our study provides a general way to explore and understand the spontaneous superfluidity breakdown in CPTs from a single-well dispersion to a double-well one, such as, BECs in synthetic gauge fields, spin-orbit coupled BECs, and BECs in shaken optical lattices.
\end{abstract}

\maketitle
\textit{Introduction}. Engineered synthetic gauge fields for neutral atoms~\cite{D.Jaksch2003, Y.J.Lin2009, F.Gerbier2010, J.Dalibard2011,M.Aidelsburger2011, Kolovsky2011, J.Struck2014, N.Goldman2014} provide new opportunities to explore exotic collective quantum phenomena~\cite{L.K.Lim2008, J.Struck2012, H.Miyake2013, M.Aidelsburger2013, HuiZhai2015, Stuhl2015}.
The dispersion relation plays an important role in the emergence of many collective quantum phenomena.
Through controlling the applied external fields, the dispersion relation can be tuned from a single-well shape into a double-well one, such as, spin-orbit coupled quantum gases~\cite{Y.J.Lin2011,V.Galitski2014,S.C.Ji2015,Y.C.Zhang2016}, ultracold atoms in shaken optical lattices~\cite{C.V.Parker2013,L.C.Ha2015,L.W.Clark2016} and Bose ladders within magnetic fields~\cite{M.Atala2014,R.Wei2014,M.Piraud2015,S.Greschner2015}.
At the transition point, due to the interplay between the synthetic gauge fields and the atom-atom interactions, the Landau critical velocity vanishes~\cite{L.C.Ha2015,S.C.Ji2015,Y.C.Zhang2016} and thus the superfluid spontaneously breaks down.
Such a spontaneous superfluidity breakdown is very different from the conventional Landau instability which requires the superfluid velocity exceeding a nonzero critical velocity~\cite{Landau1941,C. Ramanr1999,R.Desbuquois2012}.
Although the static phase transitions in synthetic gauge fields have been extensively studied, the underlying dynamics of phase transitions is still unclear.

The Kibble-Zurek mechanism (KZM)~\cite{Kibble1976,Zurek1985,J.Dziarmaga2010,delCampo2014} describes the universality of real-time dynamics in continuous phase transitions.
According to the KZM, universal scalings can be derived by comparing two characteristic time scales: the reaction time and the transition time.
Not only the dynamics of thermodynamic phase transitions~\cite{T.Donner2007, C.N.Weiler2008, B.Damski2010, R.Yusupov2010,E.Witkowska2011, A.Das2012, S.W.Su2013, G.Lamporesi2013, L.Corman2014, M.Nikkhou2015, J.Sonner2015},
but also the dynamics of quantum phase transitions~\cite{Zurek2005, B.Damski2005, Damski2007, M.Uhlmann2007, J.Dziarmaga2008, C.Lee2009, J.Sabbatini2011, D.Chen2011, N.Navon2015, M.Anquez2016, L.W.Clark2016, XuJun2016} obey the KZM.
Due to their high controllability, atomic Bose-Einstein condensates (BECs) provide an excellent platform to examine KZM~\cite{T.Donner2007, C.N.Weiler2008, B.Damski2010, E.Witkowska2011, A.Das2012, S.W.Su2013, G.Lamporesi2013, L.Corman2014, M.Nikkhou2015,Damski2007,M.Uhlmann2007, J.Dziarmaga2008, C.Lee2009, J.Sabbatini2011, D.Chen2011, N.Navon2015, M.Anquez2016, L.W.Clark2016, XuJun2016}.
Usually, the spontaneous defect generation are associated with the nontrivial gapless Bogoliubov excitations (such as Higgs modes) and thus the Kibble-Zurek scalings can be given by comparing the two characteristic time scales derived from the Bogoliubov excitation gap~\cite{Damski2007,C.Lee2009,J.Sabbatini2011,XuJun2016}.
However, for atomic BECs in synthetic gauge fields~\cite{D.Jaksch2003, Y.J.Lin2009, F.Gerbier2010, M.Aidelsburger2011, Kolovsky2011, J.Struck2014, N.Goldman2014} whose dispersion relations are continuously tuned from a single-well shape to a double-well one, the spontaneous superfluidity breakdown is resulted from the spontaneous Landau instability.
Two natural questions arise: \emph{Does the spontaneous superfluidity breakdown obey the KZM? If it obeys the KZM, can one explore the universal scalings via analyzing the Landau critical velocity}?

In this Letter, we explore the universality of spontaneous superfluidity breakdown within synthetic gauge fields.
We consider an atomic BEC ladder subjected to a synthetic magnetic field, which undergoes a continuous transition from a single-well dispersion to a double-well one.
By employing a variational ansatz, we analytically give the mean-field phase diagram.
Due to the absence of nontrivial gapless Bogoliubov excitations, it is difficult to give the two characteristic time scales and the Kibble-Zurek scalings from the Bogoliubov excitation gap.
Fortunately, we find that the universal scalings can be derived from the Landau critical velocity which determines the correlation length.
In particular, the correlation length divergence and the spontaneous superfluidity breakdown are resulted from the zero Landau critical velocity at the transition point.
To extract the Kibble-Zurek scalings, we numerically simulate the real-time dynamics of the continuous transition from a single-well dispersion to a double-well one, in which the ground state changes from the Meissner phase to the broken symmetry phase.
Our numerical results show the bifurcation delay and the spontaneous vortex generation obeys the KZM.
Our study opens a new avenue for exploring and understanding the universality of spontaneous superfluidity breakdown in various systems, such as, spin-orbit coupled BECs~\cite{Y.J.Lin2011,V.Galitski2014,S.C.Ji2015,Y.C.Zhang2016}, BECs in shaken optical lattices~\cite{C.V.Parker2013,L.C.Ha2015,L.W.Clark2016}, and BECs within magnetic fields~\cite{M.Atala2014,R.Wei2014,M.Piraud2015,S.Greschner2015}.

\textit{Model and ground states}. We consider an ensemble of Bose condensed atoms in a two-leg ladder subjected to a uniform synthetic magnetic field.
The ladder potential can be created by a normal standing-wave along one direction and a bi-frequency standing-wave along the other direction~\cite{M.Atala2014}, and the uniform synthetic magnetic field can be realized by laser-induced tunneling~\cite{D.Jaksch2003, F.Gerbier2010, M.Aidelsburger2011, Kolovsky2011,N.Goldman2014, M.Aidelsburger2013, J.Struck2012,H.Miyake2013,M.Atala2014}, see Fig. 1(a).
The mean-field Hamiltonian reads as
\begin{equation}
\begin{split}
H=&-J\sum_l(\psi_{l+1,L}^*\psi_{l,L}+\psi_{l+1,R}^*\psi_{l,R}+h.c.)\\
&-K\sum_l(\psi_{l,R}^*\psi_{l,L}\text{e}^{il\phi}+h.c.)\\
&+\frac{g}{2}\sum_l(|\psi_{l,L}|^{4}+|\psi_{l,R}|^{4}).
\end{split}
\end{equation}
with $\psi_{l,\sigma}$ denoting the order parameters for the site $(l,\sigma)$ and $\sigma=\{L, R\}$ respectively labelling the \{left, right\} legs.
Here, $J$ is the intra-leg tunneling, $K\text{e}^{il\phi}$ is the spatially dependent inter-leg tunneling and $\phi$ is the magnetic flux per plaquette.
The on-site interaction $g$, which is proportional to the $s$-wave scattering length, can be tuned via Feshbach resonances~\cite{C.Chin2010}.

\begin{figure}[htb]
\includegraphics[width=1\columnwidth]{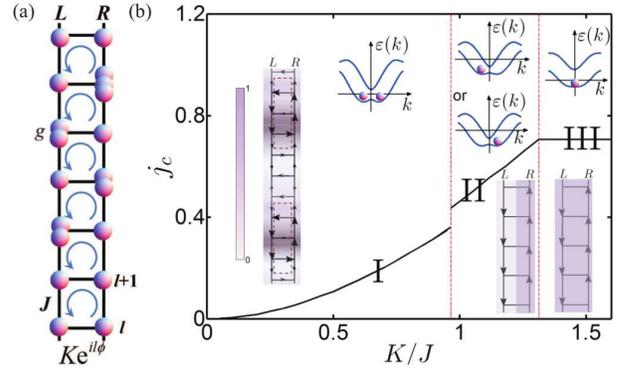}
\caption{(a) Schematic diagram for a Bose ladder in a uniform magnetic field.
Where, $J$ and $K$ are respectively the intra- and inter-leg tunneling strengthes, $g$ is the interaction strength, and $\phi$ is the magnetic flux per plaquette.
(b) The chiral current $j_c$ versus the ratio $K/J$.
There are three typical phases: (I) the vortex phase, (II) the biased ladder phase (BLP), and (III) the Meissner phase. The thickness and length of the arrows denote the current strength, which is normalized to the maximum current for the chosen $K/J$. The two dash-dotted lines label the two critical points between different phases. The dispersion relations $\varepsilon(k)$ for three typical phases are shown in insets.
The parameters are chosen as $N=5\times10^4$, $L=200$, $J=1$, $g\bar{n}/J=0.2$ and $\phi=\pi/2$.
}
\end{figure}

The determine the ground states, we implement a variational procedure based on the ansatz
\begin{equation}
\left[
\begin{array}{cc}
\psi_{l,L}\\
\psi_{l,R} \\
\end{array}
\right]
=\left[
\begin{array}{cc}
C_1 \cos\theta \text{e}^{i(k-\frac{\phi}{2})l} +C_2\sin\theta \text{e}^{-i(k+\frac{\phi}{2})l}\\
C_2 \cos\theta \text{e}^{i(k+\frac{\phi}{2})l} +C_1\sin\theta \text{e}^{-i(k-\frac{\phi}{2})l}\\
\end{array}
\right],
\end{equation}
with the total atomic number $N$, the ladder length $L$, the complex amplitudes $(C_1, C_2)$, the quasi-momentum $k$ and the angle $\theta$ ($0\le\theta\le\frac{\pi}{2}$).
By minimizing the Hamiltonian under the normalization condition $|C_1|^2+|C_2|^2=\bar{n}=N/L$, the variational parameters $(C_1, C_2, k, \theta)$ can be determined.
There are three different ground states:
(I) the vortex phase with $\{C_1\ne C_2, k \ne 0,\theta=\pi/4\}$ for $0<K/J<R^c_1$,
(II) the biased ladder phase (BLP) with $\{C_1 \ne C_2, k \ne 0,\theta=0,\pi/2\}$ for $R^c_1<K/J<R^c_2$,
and (III) the Meissner phase with $\{C_1=C_2, k = 0\}$ for $K/J>R^c_2$.
The two critical points $(R^c_1,R^c_2)$ have to be determined numerically, however, the second critical point can be analytically given as $R^c_2=\sqrt{2}-g\bar{n}/2$ for $\phi=\pi/2$.
Our numerical results show that the phase (II) disappears if there is no inter-particle interaction, that is, $R^c_1=R^c_2=\sqrt{2}$ for $g=0$.

To clarify the Meissner-like effect, we calculate the chiral current $j_c=\frac{1}{2L}\sum_{l}^{} \left<j^{||}_{l,L}-j^{||}_{l,R}\right>$, which is defined as the averaged difference between the local currents along the two legs $j^{||}_{l,\sigma}=iJ\left(\psi^*_{l+1,\sigma}\psi_{l,\sigma}-\psi^*_{l,\sigma}\psi_{l+1,\sigma}\right)$.
The chiral current increases with $K/J$ and then becomes saturated in the Meissner phase.
This is analogy to the Meissner effect in the type-II superconductor~\cite{M.Atala2014}.
The ground states and their chiral currents are consistent with the variational prediction~\cite{R.Wei2014}, see Fig. 1(b).

The phase transitions are characterized by the changes in the band structure (dispersion relation) and the chiral current.
The transition between the vortex phase and the BLP is a first-order phase transition, which corresponds to a jump in the chiral current $j_c$ and the change from complete to single occupancy of the two band-minima at finite quasi-momentum $\pm k$.
The transition between the Meissner phase and the BLP is a continuous phase transition, in which the chiral current $j_c$ is continuous and the lowest band continuously changes from a single-well shape to a double-well one.
Below we concentrate on discussing the continuous phase transition.

\textit{Landau critical velocity and correlation length}. In a continuous phase transition, the critical slowing down can be understood by the divergence of either the relaxation time $\tau_{r}$ or the correlation length $\xi$. For our system, it is difficult to give the relaxation time $\tau_{r}$. Below, we show how to derive the universal scaling of $\xi$ from the Landau critical velocity,
\begin{equation}
v_L=\mathop{\text{min}}\limits_q\left|{\omega}/{q}\right|.
\end{equation}
Here the Bogoliubov excitation gap $\omega$ is a function of the quasi-momentum $q$.
According to the Landau criterion, if the superfluid velocity exceeds $v_L$, elementary excitations appear due to the conservation of energy and momentum.
This indicates that elementary excitation will take place spontaneously if $v_L=0$.

To give $v_L$, we implement the Bogoliubov analysis to obtain the excitation gap $\omega$. We express the perturbed ground-state as,
\begin{equation}
\left[
\begin{array}{cc}
\psi_{l,L}(t)\\
\psi_{l,R}(t)\\
\end{array}
\right]
=\left[
\begin{array}{cc}
\left(C_1+\delta\psi_{l,L}(t)\right)\text{e}^{i(k-\frac{\phi}{2})l}\\
\left(C_2+\delta\psi_{l,R}(t)\right)\text{e}^{i(k+\frac{\phi}{2})l}\\
\end{array}
\right]
\text{e}^{-i\mu t},
\end{equation}
with the perturbation terms
\begin{equation}
\delta\psi_{l,\sigma}(t)=\sum_q u_{q,\sigma}\text{e}^{i(ql-\omega t)}+v_{q,\sigma}^*\text{e}^{-i(ql-\omega t)}.
\end{equation}
Here, the discrete quasi-momenta are given as $q=\frac{2m\pi}{L}$ with the integers $m=\left\{-L/2,-L/2+1,\cdots, L/2-1\right\}$,
and the perturbation amplitudes $\left(u_{q,\sigma}, v_{q,\sigma}\right)$ are complex numbers.
Inserting the perturbed state into the time-evolution equation, $i\partial \psi_{l,\sigma}/\partial t = \partial H/\partial \psi_{l,\sigma}^{*}$, one can obtain the Bogoliubov-de Gennes (BdG) equation.
Therefore, $\omega$ can be obtained by diagonalizing the BdG equation.
At the critical point, $K=K_c=J(\sqrt{2}-g\bar{n}/2)$, the low-energy long-wavelength excitation behaves as
\begin{equation}
\lim_{q \rightarrow 0} \omega(q) \approx q^2\sqrt{\frac{g\bar{n}}{2\sqrt2}}.
\end{equation}
As $\omega(q) \propto |q|^z$ when $q \rightarrow 0$~\cite{Sachdev2011,Robinson2011,A.Polkovnikov2011}, we have the dynamical critical exponent $z=2$.

\begin{figure}[htb]
\includegraphics[width=1\columnwidth]{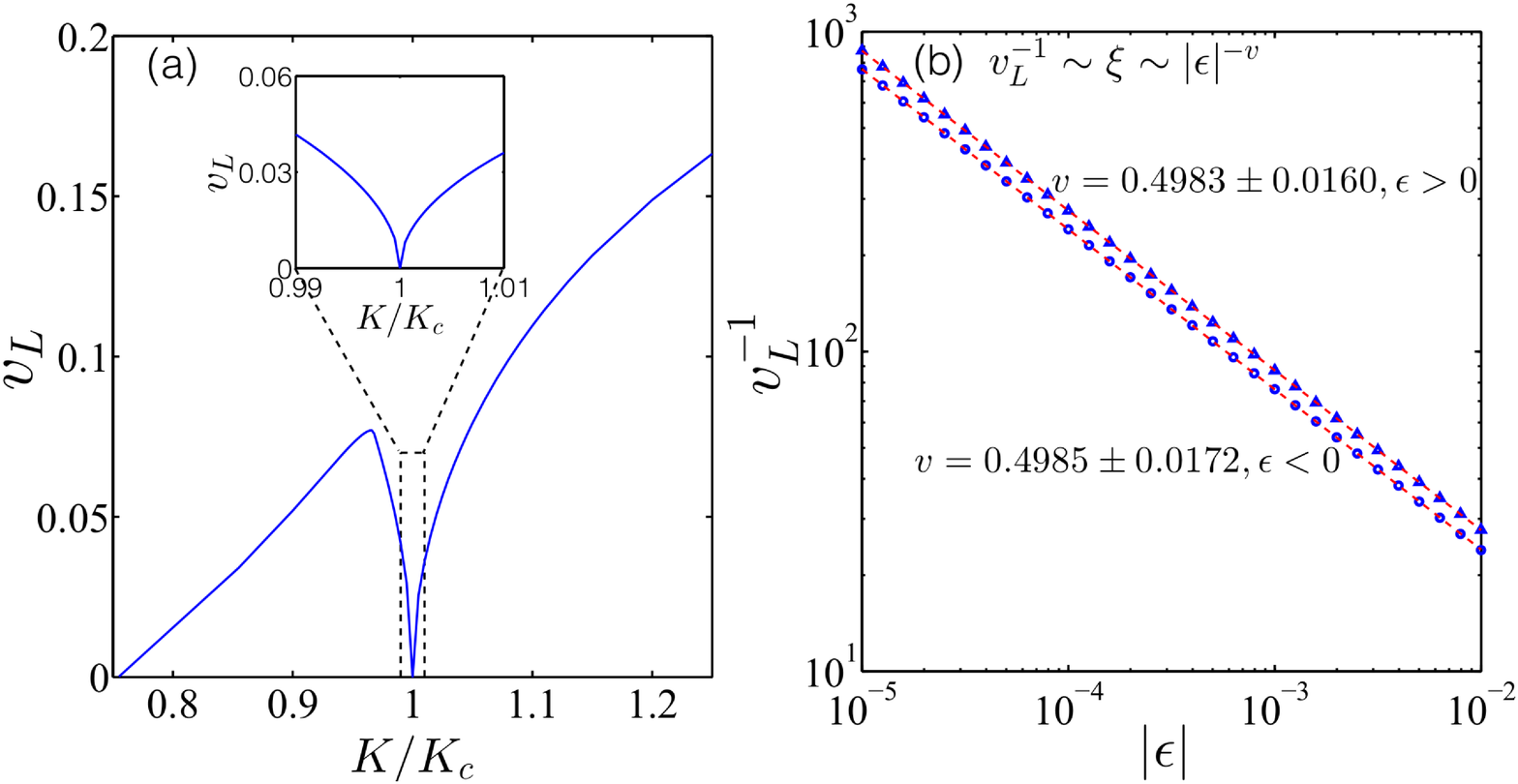}
  \caption{(a) The Landau critical velocity $v_L$ versus $K/K_c$.
  Inset: the critical region of the Landau critical velocity.
  (b) The inverse critical velocity $v_L^{-1}$ versus $\left|\epsilon\right|=\left|K(t)-K_c\right|/K_c$.
  The open triangles and circles correspond to the Meissner phase and the biased ladder phase, respectively.
  Here, to minimize finite-size effects, we choose $L=50000$ and the other parameters as same as the ones for Fig.~1.}
\end{figure}

In Fig. 2~(a), we show the dependence of $v_L$ on $K/K_c$.
The Landau critical velocity $v_L$ gradually decreases to zero when $K \rightarrow K_c$.
The notch of $v_L$ near $K=K_c$ originates from the softening phonon mode $\omega(q)\propto q^2$, which implies $|\omega(q)/q|\to 0$ as $q\to 0$~\cite{S.C.Ji2015}.
In particular, the vanishing critical velocity at the critical point will result the spontaneous superfluidity breakdown and the spontaneous elementary excitations.

Usually, the correlation length $\xi$ is defined by the equality between the kinetic energy per particle $\hbar^2/(2m\xi^2)$ and the interaction energy per particle $g\bar{n}$.
However, the Landau critical velocity $v_L$ provides another general definition according to $\xi=\hbar/(mv_L)$, which is consistent with the usual definition~\cite{S.Giorgini2008}.
This means that, near the critical point, the inverse critical velocity $v_L^{-1}$ scales as
\begin{equation}
v_L^{-1}\sim\xi\sim|\epsilon|^{-\nu}.
\end{equation}
In Fig.~2(b), we show the dependence of $v_L^{-1}$ on $\left|\epsilon\right|=\left|K(t)-K_c\right|/K_c$.
Our numerical results show $v_L^{-1} \sim |\epsilon|^{-b}$ with $b=0.4985 \pm 0.0172$ and $0.4983 \pm 0.0160$ for the BLP and the Meissner phase, respectively.
This indicates that the static correlation-length critical exponent $\nu=1/2$.

\textit{Kibble-Zurek scalings}. Now we discuss the universal scalings of the real-time dynamics across the critical point.
To drive the system from the Meissner phase to the BLP, we quench the tunneling strength $K$ according to
\begin{equation}
K(t)=K_c(1-t/\tau_Q)
\end{equation}
where $\tau_Q$ is the quench time.
In the vicinity of the critical point, both the relaxation time $\tau_{r}$ and the correlation length $\xi$ diverge as
\begin{equation}
\tau_{r} \sim |\epsilon|^{-z\nu},~~~\xi\sim |\epsilon|^{-v},
\end{equation}
where $\epsilon(t)=\left[K(t)-K_c\right]/K_c$ is the dimensionless distance from the critical point and $(z, \nu)$ are the critical exponents.
Due to the critical slowing down caused by the divergent relaxation time, a system driven across its critical point has no sufficient time to follow its instantaneous ground state no matter how slow it is driven.

The Kibble-Zurek scalings can be derived by comparing the relaxation time $\tau_r$ and the transition time $\tau_t=|\epsilon/\dot{\epsilon}|$ (where $\dot{\epsilon}=d\epsilon/dt$). The system evolves adiabatically if $\tau_r<\tau_t$, otherwise the adiabaticity breaks down.
Defining the freezing time $\hat{t}$ with $\tau_r(\hat{t})=\tau_t(\hat{t})=|\hat{t}|$, the two characteristic times change from $\tau_r<\tau_t$ to $\tau_r>\tau_t$ when the time $t$ changes from $|t|>|\hat{t}|$ to $|t|<|\hat{t}|$.
Thus, at the freezing time $\hat{t}$, the dimensionless distance $\epsilon(\hat{t})$ and the correlation length $\xi(\hat{t})$ exhibit universal power laws,
\begin{equation}
\hat{\epsilon}=\epsilon(\hat{t})\sim\tau_Q^{-1/(1+z\nu)},
~~~\hat{\xi}=\xi(\hat{t})\sim\tau_Q^{\nu/(1+z\nu)},
\end{equation}
with respect to $\tau_Q$.
After the system is driven through the critical point, distant parts of the system choose to break the symmetry randomly
and defects (discrete vortices) are spontaneously created.
The total number of generated vortices scales as
\begin{equation}
N_v \sim \hat{\xi}^{-d}\sim\tau_Q^{-d\nu/(1+z\nu)},
\end{equation}
with $d$ the dimension of the system.

\textit{Numerical scalings}. Below we show how to numerically extract the Kibble-Zurek scalings from the real-time dynamics.
According to the time-evolution equation $i\partial \psi_{l,\sigma}/\partial t = \partial H/\partial \psi_{l,\sigma}^{*}$, we simulate the quenching process [$K(t)=K_c(1-t/\tau_Q)$] for different $\tau_Q$.
When the time increases from $t<0$ to $t>0$, the system goes from the Meissner phase into the BLP.
In our simulation, the parameters are chosen as $N=5\times10^4$, $L=200$, $J=1$, $\phi=\pi/2$ and $g\bar{n}=0.2$.
For each $\tau_Q$, we perform 150 runs of simulations under random initial fluctuations.
In a single run, we calculate the bifurcation delay $b_d = |K(\hat{t})-K_c|$ and the vortex number $N_v$ and then give their averaged values for each $\tau_Q$.

In Fig.~3, we show the universal scaling of the averaged bifurcation delay.
As a signature of the impulse regime resulted from the critical slowing down, the dynamic chiral current $j_c(t)$ keep unchanged in the duration of $-\hat{t}<t<\hat{t}$.
Unlike the static case, $j_c(t)$ does not decrease immediately after the critical point $t=0$ but starts to decrease after the impulse-adiabatic transition at $t=\hat{t}$, see the inset of Fig.~3.
In a single run, $\hat{t}$ is numerically determined by $\delta j_c=\left|(j_c(\hat{t})-j_c^{\text{max}})/j_c^{\text{max}}\right|=0.005$ with the maximum chiral current $j_c^{\text{max}}$, and the bifurcation delay is given as $b_d=\left|K(\hat{t})-K_c\right| \propto \left|\hat{\epsilon}\right|$.
Our numerical results show that the averaged bifurcation delay scales as $\overline{b_d} = \overline{\left|K(\hat{t})-K_c\right|} \sim\tau_Q^{-0.5026 \pm 0.0040}$, which agrees with the analytical scaling $b_d \propto \left|\hat{\epsilon}\right| \sim\tau_Q^{-1/\left(1+z\nu\right)}$ with $1/\left(1+z\nu\right)=1/2$.

\begin{figure}[htb]
\includegraphics[width=1\columnwidth]{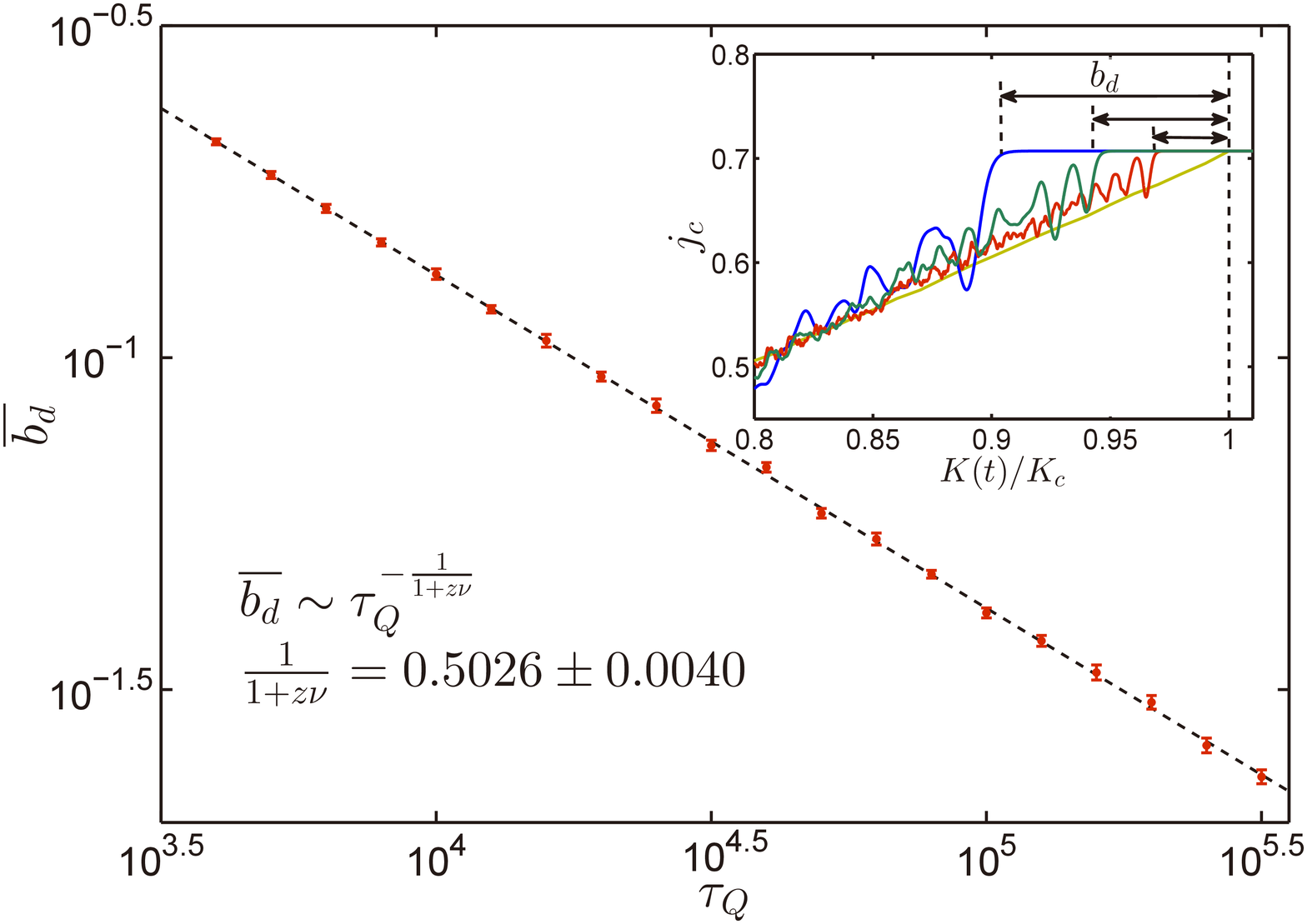}
\caption{The universal scaling of the averaged bifurcation delay $\overline{b_d}$.
  The error bars denote the standard deviation.
  Inset: the chiral current $j_c$ versus $K(t)/K_c$ for different $\tau_Q$.
  All parameters are chosen as same as the ones for Fig.~1.}
\end{figure}

In Fig.~4, we show the universal scaling of the averaged vortex number.
In a single run, we count the number of discrete vortices at a certain time $t_v$ after the impulse-adiabatic transition at $t=\hat{t}$.
As $\hat{t}$ is defined as the time where $\delta j_c = 0.005$, we count the vortex numbers at different $t_v$ where $0.005\le\delta j_c\le0.02$ and find similar scalings of the averaged vortex numbers with respect to $\tau_Q$.
The counting of vortex numbers is proceeded by analyzing the current patterns.
To minimize numerical errors, we take the inter-leg current $j_{\bot}$ to be zero if $|j_{\bot}/j_{\bot}^{\text{max}}|$ is less than a threshold 0.01.
We checked that the results are unaffected for other reasonable thresholds between $0.001$ and $0.05$.
Our numerical results show that the averaged vortex numbers follows $\overline{N_v}\sim\tau_Q^{-0.2510 \pm 0.0387}$, which is consistent with the analytical scaling $N_v\sim\tau_Q^{-d\nu/(1+z\nu)}$ with $d\nu/(1+z\nu)=1/4$.
%
%We notice that the vortex number flattens for long quench times, which are likely due to unavoidable miscountings of fluctuations as vortices. (??Is there any similar behavior in existed Refs??)

\begin{figure}[htb]
\includegraphics[width=1\columnwidth]{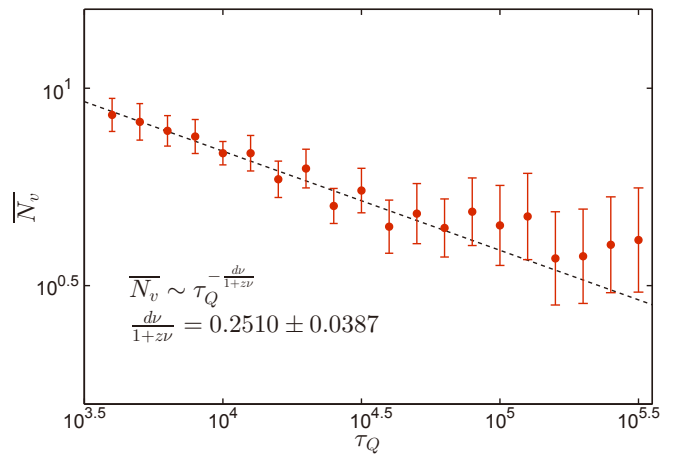}
  \caption{The universal scaling of the averaged vortex number $\overline{N_v}$ counted at the time where $\delta j_c=\left|(j_c(t)-j_c^{\text{max}})/j_c^{\text{max}}\right| =0.005$.
  The error bars denote the standard deviation.
  All parameters are chosen as same as the ones for Fig.1.}
\end{figure}

Combining the scalings for the inverse critical velocity $v_L^{-1}$, the averaged bifurcation delay $\overline{b_d}$ and the averaged vortex number $\overline{N_v}$, one may determine the dimension $d$ and the critical exponents $(z,\nu)$.
From the numerical results shown in Figs. 2(b), 3 and 4, we give the dimension $d=1$, the static correlation-length critical exponent $\nu=1/2$ and the dynamic critical exponent $z=2$.
Actually, $z=2$ is a direct result of the unique quartic dispersion at the critical point.
These critical exponents are in the same universal class for the continuous phase transitions, in which the dispersion continuously changes from a single-well shape to a double-well one~\cite{Y.J.Lin2011, V.Galitski2014, S.C.Ji2015, Y.C.Zhang2016, C.V.Parker2013, L.C.Ha2015, L.W.Clark2016, M.Atala2014, R.Wei2014, M.Piraud2015, S.Greschner2015}.

\textit{Summary and discussion}. To summarize, we reveal the universality of spontaneous superfluidity breakdown within synthetic gauge fields.
We find that the spontaneous superfluidity breakdown obeys the KZM and extract the critical exponents from the Landau critical velocity and the critical spatial-temporal dynamics.
The numerical scalings extracted from the critical spatial-temporal dynamics well agree with the analytical Kibble-Zurek scalings.
Our study provide a general approach to explore and understand the dynamic universality of continuous phase transitions involving continuous variation from a single-well dispersion to a double-well one~\cite{Y.J.Lin2011, V.Galitski2014, S.C.Ji2015, Y.C.Zhang2016, C.V.Parker2013, L.C.Ha2015, L.W.Clark2016, M.Atala2014, R.Wei2014, M.Piraud2015, S.Greschner2015}.
As an experimental evidence, the critical spatial-temporal dynamics of BECs in a shaken optical lattice~\cite{L.W.Clark2016} share the same critical exponents $(z=2,\nu=1/2)$ for ours.

At last, we briefly discuss the experimental feasibility.
Based upon the recent experiment~\cite{M.Atala2014}, our system can be realized and the considered phase transition can be observed if the interaction is sufficiently strong.
The interaction strength can be enhanced via Feshbach resonance~\cite{C.Chin2010} or tuning the transverse confinement~\cite{L.C.Ha2013}.
To drive the system across the considered phase transition, one may gradually decrease the inter-leg potential barriers via decreasing the laser intensity.
The Kibble-Zurek scalings can be extracted by counting the vortex number and measuring the chiral current via the well-developed high-resolution imaging~\cite{W.S.Bakr2010,J.F.Sherson2010}.

\acknowledgments{
%We thank ... for valuable discussions.
%
This work was supported by the National Basic Research Program of China (Grant No. 2012CB821305) and the National Natural Science Foundation of China (Grants No. 11374375 and No. 11574405).
}

%\end{CJK*}

\begin{thebibliography}{99}
\expandafter\ifx\csname natexlab\endcsname\relax\def\natexlab#1{#1}\fi
\expandafter\ifx\csname bibnamefont\endcsname\relax
  \def\bibnamefont#1{#1}\fi
\expandafter\ifx\csname bibfnamefont\endcsname\relax
  \def\bibfnamefont#1{#1}\fi
\expandafter\ifx\csname citenamefont\endcsname\relax
  \def\citenamefont#1{#1}\fi
\expandafter\ifx\csname url\endcsname\relax
  \def\url#1{\texttt{#1}}\fi
\expandafter\ifx\csname urlprefix\endcsname\relax\def\urlprefix{URL }\fi
\providecommand{\bibinfo}[2]{#2}
\providecommand{\eprint}[2][]{\url{#2}}

\bibitem[{\citenamefont{Jaksch and Zoller}(2003)}]{D.Jaksch2003}
  \bibinfo{author}{\bibfnamefont{D.}~\bibnamefont{Jaksch}} \bibnamefont{and}
  \bibinfo{author}{\bibfnamefont{P.}~\bibnamefont{Zoller}},
  \bibinfo{title}{Creation of effective magnetic fields in optical lattices: the Hofstadter butterfly for cold neutral atoms},
  \bibinfo{journal}{New J. Phys.} \textbf{\bibinfo{volume}{5}},
  \bibinfo{pages}{56} (\bibinfo{year}{2003}).


\bibitem[{\citenamefont{Lin et~al.}(2009)\citenamefont{Lin, Compton,  Jim¨¦nez-Garc¨ªa, Porto, and Spielman}}]{Y.J.Lin2009}
  \bibinfo{author}{\bibfnamefont{Y.-J.} \bibnamefont{Lin}},
  \bibinfo{author}{\bibfnamefont{R.~L.} \bibnamefont{Compton}},
  \bibinfo{author}{\bibfnamefont{K.}~\bibnamefont{Jim¨¦nez-Garc¨ªa}},
  \bibinfo{author}{\bibfnamefont{J.~V.} \bibnamefont{Porto}}, \bibnamefont{and}
  \bibinfo{author}{\bibfnamefont{I.~B.} \bibnamefont{Spielman}},
  \bibinfo{title}{Synthetic magnetic fields for ultracold neutral atoms},
  \bibinfo{journal}{Nature} \textbf{\bibinfo{volume}{462}},
  \bibinfo{pages}{628} (\bibinfo{year}{2009}).

\bibitem[{\citenamefont{Gerbier and Dalibard}(2010)}]{F.Gerbier2010}
  \bibinfo{author}{\bibfnamefont{F.}~\bibnamefont{Gerbier}} \bibnamefont{and}
  \bibinfo{author}{\bibfnamefont{J.}~\bibnamefont{Dalibard}},
  \bibinfo{title}{Gauge fields for ultracold atoms in optical superlattices},
  \bibinfo{journal}{New J. Phys.} \textbf{\bibinfo{volume}{12}},
  \bibinfo{pages}{033007} (\bibinfo{year}{2010}).

\bibitem[{\citenamefont{Dalibard et.al}(2011)}]{J.Dalibard2011}
  \bibinfo{author}{\bibfnamefont{J.}~\bibnamefont{Dalibard}},
  \bibinfo{author}{\bibfnamefont{F.}~\bibnamefont{Gerbier}},
  \bibinfo{author}{\bibfnamefont{G.}~\bibnamefont{Juzeli¨±nas}}, \bibnamefont{and}
  \bibinfo{author}{\bibfnamefont{P.}~\bibnamefont{$\ddot{\text{O}}$hberg}},
  \bibinfo{title}{\emph{Colloquium}: Artificial gauge potentials for neutral atoms},
  \bibinfo{journal}{Rev. Mod. Phys. } \textbf{\bibinfo{volume}{83}},
  \bibinfo{pages}{1523} (\bibinfo{year}{2011}).


\bibitem[{\citenamefont{Kolovsky}(2011)}]{Kolovsky2011}
  \bibinfo{author}{\bibfnamefont{A.~R.} \bibnamefont{Kolovsky}},
  \bibinfo{title}{Creating artificial magnetic fields for cold atoms by photon-assisted tunneling},
  \bibinfo{journal}{EPL} \textbf{\bibinfo{volume}{93}},
  \bibinfo{pages}{20003} (\bibinfo{year}{2011}).

\bibitem[{\citenamefont{Aidelsburger et~al.}(2011)\citenamefont{Aidelsburger,  Atala, Nascimb\`ene, Trotzky, Chen, and Bloch}}]{M.Aidelsburger2011}
  \bibinfo{author}{\bibfnamefont{M.}~\bibnamefont{Aidelsburger}},
  \bibinfo{author}{\bibfnamefont{M.}~\bibnamefont{Atala}},
  \bibinfo{author}{\bibfnamefont{S.}~\bibnamefont{Nascimb\`ene}},
  \bibinfo{author}{\bibfnamefont{S.}~\bibnamefont{Trotzky}},
  \bibinfo{author}{\bibfnamefont{Y.-A.} \bibnamefont{Chen}}, \bibnamefont{and}
  \bibinfo{author}{\bibfnamefont{I.}~\bibnamefont{Bloch}},
  \bibinfo{title}{Experimental Realization of Strong Effective Magnetic Fields in an Optical Lattice},
  \bibinfo{journal}{Phys. Rev. Lett.} \textbf{\bibinfo{volume}{107}},
  \bibinfo{pages}{255301} (\bibinfo{year}{2011}).

\bibitem[{\citenamefont{Struck et~al.}(2013)\citenamefont{Struck, Weinberg,  $\ddot{\text{O}}$lschl$\ddot{\text{a}}$ger, Windpassinger, Simonet,  Sengstock, H$\ddot{\text{a}}$ppner, Hauke, Eckardt, Lewenstein  et~al.}}]{J.Struck2014}
  \bibinfo{author}{\bibfnamefont{J.}~\bibnamefont{Struck}},
  \bibinfo{author}{\bibfnamefont{M.}~\bibnamefont{Weinberg}},
  \bibinfo{author}{\bibfnamefont{C.}~\bibnamefont{$\ddot{\text{O}}$lschl$\ddot{\text{a}}$ger}},
  \bibinfo{author}{\bibfnamefont{P.}~\bibnamefont{Windpassinger}},
  \bibinfo{author}{\bibfnamefont{J.}~\bibnamefont{Simonet}},
  \bibinfo{author}{\bibfnamefont{K.}~\bibnamefont{Sengstock}},
  \bibinfo{author}{\bibfnamefont{R.}~\bibnamefont{H$\ddot{\text{a}}$ppner}},
  \bibinfo{author}{\bibfnamefont{P.}~\bibnamefont{Hauke}},
  \bibinfo{author}{\bibfnamefont{A.}~\bibnamefont{Eckardt}},
  \bibinfo{author}{\bibfnamefont{M.}~\bibnamefont{Lewenstein}},
  \bibinfo{title}{Engineering Ising-XY spin-models in a triangular lattice using tunable artificial gauge fields},
  \bibinfo{journal}{Nat. Phys.}
  \textbf{\bibinfo{volume}{9}}, \bibinfo{pages}{738}
  (\bibinfo{year}{2013}).

\bibitem[{\citenamefont{Goldman}(2014)}]{N.Goldman2014}
  \bibinfo{author}{\bibfnamefont{N.} \bibnamefont{Goldman}},
  \bibinfo{author}{\bibfnamefont{G.} \bibnamefont{Juzeli¨±nas}},
  \bibinfo{author}{\bibfnamefont{P.} \bibnamefont{\"Ohber}}, \bibnamefont{and}
  \bibinfo{author}{\bibfnamefont{I.~B.}~\bibnamefont{Spielman}},
  \bibinfo{title}{Light-induced gauge fields for ultracold atoms},
  \bibinfo{journal}{Rep. Prog. Phys.} \textbf{\bibinfo{volume}{77}},
  \bibinfo{pages}{126401} (\bibinfo{year}{2014}).

\bibitem[{\citenamefont{Aidelsburger et~al.}(2013)\citenamefont{Aidelsburger,  Atala, Lohse, Barreiro, Paredes, and Bloch}}]{M.Aidelsburger2013}
  \bibinfo{author}{\bibfnamefont{M.}~\bibnamefont{Aidelsburger}},
  \bibinfo{author}{\bibfnamefont{M.}~\bibnamefont{Atala}},
  \bibinfo{author}{\bibfnamefont{M.}~\bibnamefont{Lohse}},
  \bibinfo{author}{\bibfnamefont{J.~T.} \bibnamefont{Barreiro}},
  \bibinfo{author}{\bibfnamefont{B.}~\bibnamefont{Paredes}}, \bibnamefont{and}
  \bibinfo{author}{\bibfnamefont{I.}~\bibnamefont{Bloch}},
  \bibinfo{title}{Realization of the Hofstadter Hamiltonian with Ultracold Atoms in Optical Lattices},
  \bibinfo{journal}{Phys. Rev. Lett.} \textbf{\bibinfo{volume}{111}},
  \bibinfo{pages}{185301} (\bibinfo{year}{2013}).


\bibitem[{\citenamefont{Lim et~al.}(2008)\citenamefont{Lim, Smith, and  Hemmerich}}]{L.K.Lim2008}
  \bibinfo{author}{\bibfnamefont{L.-K.} \bibnamefont{Lim}},
  \bibinfo{author}{\bibfnamefont{C.~M.} \bibnamefont{Smith}}, \bibnamefont{and}
  \bibinfo{author}{\bibfnamefont{A.}~\bibnamefont{Hemmerich}},
  \bibinfo{title}{Staggered-Vortex Superfluid of Ultracold Bosons in an Optical Lattice},
  \bibinfo{journal}{Phys. Rev. Lett.} \textbf{\bibinfo{volume}{100}},
  \bibinfo{pages}{130402} (\bibinfo{year}{2008}).

\bibitem[{\citenamefont{Struck et~al.}(2012)\citenamefont{Struck, \"Olschl\"ager, Weinberg, Hauke, Simonet, Eckardt, Lewenstein, Sengstock, and  Windpassinger}}]{J.Struck2012}
  \bibinfo{author}{\bibfnamefont{J.}~\bibnamefont{Struck}},
  \bibinfo{author}{\bibfnamefont{C.}~\bibnamefont{\"Olschl\"ager}},
  \bibinfo{author}{\bibfnamefont{M.}~\bibnamefont{Weinberg}},
  \bibinfo{author}{\bibfnamefont{P.}~\bibnamefont{Hauke}},
  \bibinfo{author}{\bibfnamefont{J.}~\bibnamefont{Simonet}},
  \bibinfo{author}{\bibfnamefont{A.}~\bibnamefont{Eckardt}},
  \bibinfo{author}{\bibfnamefont{M.}~\bibnamefont{Lewenstein}},
  \bibinfo{author}{\bibfnamefont{K.}~\bibnamefont{Sengstock}},
  \bibnamefont{and}
  \bibinfo{author}{\bibfnamefont{P.}~\bibnamefont{Windpassinger}},
  \bibinfo{title}{Tunable Gauge Potential for Neutral and Spinless Particles in Driven Optical Lattices},
  \bibinfo{journal}{Phys. Rev. Lett.} \textbf{\bibinfo{volume}{108}},
  \bibinfo{pages}{225304} (\bibinfo{year}{2012}).

\bibitem[{\citenamefont{Miyake et~al.}(2013)\citenamefont{Miyake, Siviloglou,  Kennedy, Burton, and Ketterle}}]{H.Miyake2013}
  \bibinfo{author}{\bibfnamefont{H.}~\bibnamefont{Miyake}},
  \bibinfo{author}{\bibfnamefont{G.~A.} \bibnamefont{Siviloglou}},
  \bibinfo{author}{\bibfnamefont{C.~J.} \bibnamefont{Kennedy}},
  \bibinfo{author}{\bibfnamefont{W.~C.} \bibnamefont{Burton}},
  \bibnamefont{and} \bibinfo{author}{\bibfnamefont{W.}~\bibnamefont{Ketterle}},
  \bibinfo{title}{Realizing the Harper Hamiltonian with Laser-Assisted Tunneling in Optical Lattices},
  \bibinfo{journal}{Phys. Rev. Lett.} \textbf{\bibinfo{volume}{111}},
  \bibinfo{pages}{185302} (\bibinfo{year}{2013}).

\bibitem[{\citenamefont{Zhai}(2015)}]{HuiZhai2015}
  \bibinfo{author}{\bibfnamefont{H.} \bibnamefont{Zhai}},
  \bibinfo{title}{Degenerate quantum gases with spin-orbit coupling: a review},
  \bibinfo{journal}{Rep. Prog. Phys.} \textbf{\bibinfo{volume}{78}},
  \bibinfo{pages}{026001} (\bibinfo{year}{2015}).

\bibitem{Stuhl2015}B. K. Stuhl, H.-I. Lu, L. M. Aycock, D. Genkina, and I. B. Spielman, Visualizing edge states with an atomic Bose gas in the quantum Hall regime, Science \textbf{349}, 1514 (2015).

\bibitem[{\citenamefont{Lin et~al.}(2011)\citenamefont{Lin, Jimenez-Garcia, and Spielman}}]{Y.J.Lin2011}
  \bibinfo{author}{\bibfnamefont{Y.~J.} \bibnamefont{Lin}},
  \bibinfo{author}{\bibfnamefont{K.}~\bibnamefont{Jimenez-Garcia}},
  \bibnamefont{and} \bibinfo{author}{\bibfnamefont{I.~B.}
  \bibnamefont{Spielman}},
  \bibinfo{title}{Spin-orbit-coupled bose-einstein condensates},
  \bibinfo{journal}{Nature}
  \textbf{\bibinfo{volume}{471}}, \bibinfo{pages}{83} (\bibinfo{year}{2011}).

\bibitem[{\citenamefont{Galitski and Spielman}(2014)}]{V.Galitski2014}
  \bibinfo{author}{\bibfnamefont{V.}~\bibnamefont{Galitski}} \bibnamefont{and}
  \bibinfo{author}{\bibfnamefont{I.~B.} \bibnamefont{Spielman}},
  \bibinfo{title}{Spin-orbit coupling in quantum gases},
  \bibinfo{journal}{Nature} \textbf{\bibinfo{volume}{494}}, \bibinfo{pages}{49}
  (\bibinfo{year}{2014}).

\bibitem[{\citenamefont{Ji et~al.}(2015)\citenamefont{Pan}}]{S.C.Ji2015}
  \bibinfo{author}{\bibfnamefont{S.-C.} \bibnamefont{Ji}},
  \bibinfo{author}{\bibfnamefont{L.}~\bibnamefont{Zhang}},
  \bibinfo{author}{\bibfnamefont{X.-T}~\bibnamefont{Xu}},
  \bibinfo{author}{\bibfnamefont{W.}~\bibnamefont{Zhan}},
  \bibinfo{author}{\bibfnamefont{Y.~J}~\bibnamefont{Deng}},
  \bibinfo{author}{\bibfnamefont{S.} \bibnamefont{Chen}}, \bibnamefont{and}
  \bibinfo{author}{\bibfnamefont{J.-W}~\bibnamefont{Pan}},
  \bibinfo{title}{Softening of Roton and Phonon Modes in a Bose-Einstein Condensate with Spin-Orbit Coupling},
  \bibinfo{journal}{Phys. Rev. Lett.} \textbf{\bibinfo{volume}{114}},
  \bibinfo{pages}{105301} (\bibinfo{year}{2015}).

\bibitem[{\citenamefont{Zhang et~al.}(2016)\citenamefont{Zhang, Yu, Ng, Pitaevskii, and Stringari}}]{Y.C.Zhang2016}
  \bibinfo{author}{\bibfnamefont{Y.-C.} \bibnamefont{Zhang}},
  \bibinfo{author}{\bibfnamefont{Z.-Q.}~\bibnamefont{Yu}},
  \bibinfo{author}{\bibfnamefont{T.~K.}~\bibnamefont{Ng}},
  \bibinfo{author}{\bibfnamefont{L.}~\bibnamefont{Pitaevskii}}, \bibnamefont{and}
  \bibinfo{author}{\bibfnamefont{S.}~\bibnamefont{Stringari}},
  \bibinfo{title}{Superfluid Density of a Spin-orbit Coupled Bose Gas},
  \bibinfo{journal}{arXiv:1605.02136} (\bibinfo{year}{2016}).

\bibitem[{\citenamefont{Parker et~al.}(2013)\citenamefont{Parker, Ha, and Chin}}]{C.V.Parker2013}
  \bibinfo{author}{\bibfnamefont{C.~V.} \bibnamefont{Parker}},
  \bibinfo{author}{\bibfnamefont{L.-C.} \bibnamefont{Ha}}, \bibnamefont{and}
  \bibinfo{author}{\bibfnamefont{C.}~\bibnamefont{Chin}},
  \bibinfo{title}{Direct observation of effective ferromagnetic domains of cold atoms in a shaken optical lattice},
  \bibinfo{journal}{Nat. Phys.} \textbf{\bibinfo{volume}{9}},
  \bibinfo{pages}{769} (\bibinfo{year}{2013}).

\bibitem[{\citenamefont{Ha et~al.}(2015)\citenamefont{Ha, Clark,Parker, Anderson, and  Chin}}]{L.C.Ha2015}
  \bibinfo{author}{\bibfnamefont{L.-C.} \bibnamefont{Ha}},
  \bibinfo{author}{\bibfnamefont{L.~W.} \bibnamefont{Clark}},
  \bibinfo{author}{\bibfnamefont{C.~V.} \bibnamefont{Parker}},
  \bibinfo{author}{\bibfnamefont{B.~M. } \bibnamefont{Anderson}}, \bibnamefont{and}
  \bibinfo{author}{\bibfnamefont{C.}~\bibnamefont{Chin}},
  \bibinfo{title}{Roton-Maxon Excitation Spectrum of Bose Condensates in a Shaken Optical Lattice},
  \bibinfo{journal}{Phys. Rev. Lett.} \textbf{\bibinfo{volume}{114}},
  \bibinfo{pages}{055301 } (\bibinfo{year}{2015}).


\bibitem[{\citenamefont{Clark et~al.}(2016)\citenamefont{Clark, Feng, and  Chin}}]{L.W.Clark2016}
  \bibinfo{author}{\bibfnamefont{L.~W.} \bibnamefont{Clark}},
  \bibinfo{author}{\bibfnamefont{L.} \bibnamefont{Feng}}, \bibnamefont{and}
  \bibinfo{author}{\bibfnamefont{C.}~\bibnamefont{Chin}},
  \bibinfo{title}{Universal space-time scaling symmetry in the dynamics of bosons across a quantum phase transition},
  \bibinfo{journal}{arXiv:1605.01023}.

\bibitem[{\citenamefont{Atala et~al.}(2014)\citenamefont{Atala, Aidelsburger, Lohse, Barreiro, Paredes, and Bloch}}]{M.Atala2014}
  \bibinfo{author}{\bibfnamefont{M.}~\bibnamefont{Atala}},
  \bibinfo{author}{\bibfnamefont{M.}~\bibnamefont{Aidelsburger}},
  \bibinfo{author}{\bibfnamefont{M.}~\bibnamefont{Lohse}},
  \bibinfo{author}{\bibfnamefont{J.~T.} \bibnamefont{Barreiro}},
  \bibinfo{author}{\bibfnamefont{B.}~\bibnamefont{Paredes}}, \bibnamefont{and}
  \bibinfo{author}{\bibfnamefont{I.}~\bibnamefont{Bloch}},
  \bibinfo{title}{Observation of chiral currents with ultracold atoms in bosonic ladders},
  \bibinfo{journal}{Nat. Phys.} \textbf{\bibinfo{volume}{10}},
  \bibinfo{pages}{588} (\bibinfo{year}{2014}).


\bibitem[{\citenamefont{Wei and Mueller}(2014)}]{R.Wei2014}
  \bibinfo{author}{\bibfnamefont{R.}~\bibnamefont{Wei}} \bibnamefont{and}
  \bibinfo{author}{\bibfnamefont{E.~J.} \bibnamefont{Mueller}},
  \bibinfo{title}{Theory of bosons in two-leg ladders with large magnetic fields},
  \bibinfo{journal}{Phys. Rev. A} \textbf{\bibinfo{volume}{89}},
  \bibinfo{pages}{063617} (\bibinfo{year}{2014}).

\bibitem[{\citenamefont{Piraud et~al.}(2015)\citenamefont{Piraud, Heidrich-Meisner, McCulloch, Greschner, Vekua, and Schollw\"ock}}]{M.Piraud2015}
  \bibinfo{author}{\bibfnamefont{M.}~\bibnamefont{Piraud}},
  \bibinfo{author}{\bibfnamefont{F.}~\bibnamefont{Heidrich-Meisner}},
  \bibinfo{author}{\bibfnamefont{I.~P.} \bibnamefont{McCulloch}},
  \bibinfo{author}{\bibfnamefont{S.}~\bibnamefont{Greschner}},
  \bibinfo{author}{\bibfnamefont{T.}~\bibnamefont{Vekua}}, \bibnamefont{and}
  \bibinfo{author}{\bibfnamefont{U.}~\bibnamefont{Schollw\"ock}},
  \bibinfo{title}{Vortex and Meissner phases of strongly interacting bosons on a two-leg ladder},
  \bibinfo{journal}{Phys. Rev. B} \textbf{\bibinfo{volume}{91}},
  \bibinfo{pages}{140406} (\bibinfo{year}{2015}).

\bibitem[{\citenamefont{Greschner et~al.}(2015)\citenamefont{Greschner, Piraud, Heidrich-Meisner, McCulloch, Schollw\"ock, and Vekua}}]{S.Greschner2015}
  \bibinfo{author}{\bibfnamefont{S.}~\bibnamefont{Greschner}},
  \bibinfo{author}{\bibfnamefont{M.}~\bibnamefont{Piraud}},
  \bibinfo{author}{\bibfnamefont{F.}~\bibnamefont{Heidrich-Meisner}},
  \bibinfo{author}{\bibfnamefont{I.~P.} \bibnamefont{McCulloch}},
  \bibinfo{author}{\bibfnamefont{U.}~\bibnamefont{Schollw\"ock}},
  \bibnamefont{and} \bibinfo{author}{\bibfnamefont{T.}~\bibnamefont{Vekua}},
  \bibinfo{title}{Spontaneous Increase of Magnetic Flux and Chiral-Current Reversal in Bosonic Ladders: Swimming against the Tide},
  \bibinfo{journal}{Phys. Rev. Lett.} \textbf{\bibinfo{volume}{115}},
  \bibinfo{pages}{190402} (\bibinfo{year}{2015}).

\bibitem[{\citenamefont{Landau}(1941)}]{Landau1941}
  \bibinfo{author}{\bibfnamefont{L. ~D.} \bibnamefont{Landau}},
  \bibinfo{title}{The Theory of Superfluidity of Helium II},
  \bibinfo{journal}{J. Phys. USSR}~ \textbf{\bibinfo{volume}{5}},
  \bibinfo{pages}{71} (\bibinfo{year}{1941}).

\bibitem[{\citenamefont{Raman et~al.}(1999)}]{C. Ramanr1999}
  \bibinfo{author}{\bibfnamefont{C.}~\bibnamefont{Raman}},
  \bibinfo{author}{\bibfnamefont{M.}~\bibnamefont{K\"ohl}},
  \bibinfo{author}{\bibfnamefont{R.}~\bibnamefont{R}},
  \bibinfo{author}{\bibfnamefont{D. ~S.} \bibnamefont{D. S.}},
  \bibinfo{author}{\bibfnamefont{C.~ E.}~\bibnamefont{Kuklewic}},
  \bibinfo{author}{\bibfnamefont{Z.}~\bibnamefont{Hadzibabic}},
  \bibnamefont{and} \bibinfo{author}{\bibfnamefont{W.}~\bibnamefont{Ketterle}},
  \bibinfo{title}{Evidence for a Critical Velocity in a Bose-Einstein Condensed Gas},
  \bibinfo{journal}{Phys. Rev. Lett.} \textbf{\bibinfo{volume}{83}},
  \bibinfo{pages}{2502} (\bibinfo{year}{1999}).


\bibitem[{\citenamefont{Desbuquois et~al.}(2012)}]{R.Desbuquois2012}
  \bibinfo{author}{\bibfnamefont{R.}~\bibnamefont{Desbuquois}},
  \bibinfo{author}{\bibfnamefont{L.}~\bibnamefont{Chomaz}},
  \bibinfo{author}{\bibfnamefont{T.}~\bibnamefont{Yefsah}},
  \bibinfo{author}{\bibfnamefont{J.} \bibnamefont{L\'eonard}},
  \bibinfo{author}{\bibfnamefont{J.}~\bibnamefont{Beugnon}},
  \bibinfo{author}{\bibfnamefont{C.}~\bibnamefont{Weitenberg}},
  \bibnamefont{and} \bibinfo{author}{\bibfnamefont{J.}~\bibnamefont{Dalibard}},
  \bibinfo{title}{Superfluid behaviour of a two-dimensional Bose gas},
  \bibinfo{journal}{Nat.Phys.} \textbf{\bibinfo{volume}{8}},
  \bibinfo{pages}{645} (\bibinfo{year}{2012}).


\bibitem[{\citenamefont{Kibble}(1976)}]{Kibble1976}
  \bibinfo{author}{\bibfnamefont{T.~W.~B.} \bibnamefont{Kibble}},
  \bibinfo{title}{Topology of cosmic domains and strings},
  \bibinfo{journal}{J. Phys. A: Math. Gen.} \textbf{\bibinfo{volume}{9}},
  \bibinfo{pages}{1387} (\bibinfo{year}{1976}).

\bibitem[{\citenamefont{Zurek}(1985)}]{Zurek1985}
  \bibinfo{author}{\bibfnamefont{W.~H.} \bibnamefont{Zurek}},
  \bibinfo{title}{Cosmological experiments in superfluid helium?},
  \bibinfo{journal}{Nature} \textbf{\bibinfo{volume}{317}},
  \bibinfo{pages}{505} (\bibinfo{year}{1985}).

\bibitem[{\citenamefont{Dziarmaga}(2010)}]{J.Dziarmaga2010}
  \bibinfo{author}{\bibfnamefont{J.}~\bibnamefont{Dziarmaga}},
  \bibinfo{title}{Dynamics of a quantum phase transition and relaxation to a steady state},
  \bibinfo{journal}{Adv. Phys.} \textbf{\bibinfo{volume}{59}},
  \bibinfo{pages}{1063} (\bibinfo{year}{2010}).

\bibitem[{\citenamefont{del Campo and Zurek}(2014)}]{delCampo2014}
  \bibinfo{author}{\bibfnamefont{A.}~\bibnamefont{del Campo}} \bibnamefont{and}
  \bibinfo{author}{\bibfnamefont{W.~H.} \bibnamefont{Zurek}},
  \bibinfo{title}{Universality of phase transition dynamics: Topological defects from symmetry breaking},
  \bibinfo{journal}{Int. J. Mod. Phys. A}
  \textbf{\bibinfo{volume}{29}}, \bibinfo{pages}{1430018}
  (\bibinfo{year}{2014}).

\bibitem[{\citenamefont{Donner et~al.}(2007)\citenamefont{Donner, Ritter,  Bourdel, $\ddot{\text{O}}$ttl, K$\ddot{\text{O}}$hl, and Esslinger}}]{T.Donner2007}
  \bibinfo{author}{\bibfnamefont{T.}~\bibnamefont{Donner}},
  \bibinfo{author}{\bibfnamefont{S.}~\bibnamefont{Ritter}},
  \bibinfo{author}{\bibfnamefont{T.}~\bibnamefont{Bourdel}},
  \bibinfo{author}{\bibfnamefont{A.}~\bibnamefont{$\ddot{\text{O}}$ttl}},
  \bibinfo{author}{\bibfnamefont{M.}~\bibnamefont{K$\ddot{\text{O}}$hl}},
  \bibnamefont{and}
  \bibinfo{author}{\bibfnamefont{T.}~\bibnamefont{Esslinger}},
  \bibinfo{title}{Critical Behavior of a Trapped Interacting Bose Gas},
  \bibinfo{journal}{Science} \textbf{\bibinfo{volume}{315}},
  \bibinfo{pages}{1556} (\bibinfo{year}{2007}).

\bibitem[{\citenamefont{Weiler et~al.}(2008)\citenamefont{Weiler, Neely, Scherer, Bradley, Davis, and Anderson}}]{C.N.Weiler2008}
  \bibinfo{author}{\bibfnamefont{C.~N.} \bibnamefont{Weiler}},
  \bibinfo{author}{\bibfnamefont{T.~W.} \bibnamefont{Neely}},
  \bibinfo{author}{\bibfnamefont{D.~R.} \bibnamefont{Scherer}},
  \bibinfo{author}{\bibfnamefont{A.~S.} \bibnamefont{Bradley}},
  \bibinfo{author}{\bibfnamefont{M.~J.} \bibnamefont{Davis}}, \bibnamefont{and}
  \bibinfo{author}{\bibfnamefont{B.~P.} \bibnamefont{Anderson}},
  \bibinfo{title}{Spontaneous vortices in the formation of Bose-Einstein condensates},
  \bibinfo{journal}{Nature} \textbf{\bibinfo{volume}{455}},
  \bibinfo{pages}{948} (\bibinfo{year}{2008}).

\bibitem[{\citenamefont{Damski and Zurek}(2010)}]{B.Damski2010}
  \bibinfo{author}{\bibfnamefont{B.}~\bibnamefont{Damski}} \bibnamefont{and}
  \bibinfo{author}{\bibfnamefont{W.~H.} \bibnamefont{Zurek}},
  \bibinfo{title}{Soliton Creation During a Bose-Einstein Condensation},
  \bibinfo{journal}{Phys. Rev. Lett.} \textbf{\bibinfo{volume}{104}},
  \bibinfo{pages}{160404} (\bibinfo{year}{2010}).


\bibitem[{\citenamefont{R.Yusupov et al}(2010)}]{R.Yusupov2010}
  \bibinfo{author}{\bibfnamefont{R.}~\bibnamefont{Yusupov}} ,
  \bibinfo{author}{\bibfnamefont{T.}~\bibnamefont{Mertelj}},
  \bibinfo{author}{\bibfnamefont{V. ~V.}~\bibnamefont{Kabanov}},
  \bibinfo{author}{\bibfnamefont{S.}~\bibnamefont{Brazovskii}},
  \bibinfo{author}{\bibfnamefont{P.}~\bibnamefont{Kusar}},
  \bibinfo{author}{\bibfnamefont{J.-H.}~\bibnamefont{Chu}},
  \bibinfo{author}{\bibfnamefont{I.~R.}~\bibnamefont{Fisher}}\bibnamefont{and}
  \bibinfo{author}{\bibfnamefont{D.} \bibnamefont{Mihailovic}},
  \bibinfo{title}{Coherent dynamics of macroscopic electronic order through a symmetry breaking transition},
  \bibinfo{journal}{Nat. Phys.} \textbf{\bibinfo{volume}{6}},
  \bibinfo{pages}{681} (\bibinfo{year}{2010}).


\bibitem[{\citenamefont{Witkowska et~al.}(2011)\citenamefont{Witkowska, Deuar, Gajda, and Rzazewski}}]{E.Witkowska2011}
  \bibinfo{author}{\bibfnamefont{E.}~\bibnamefont{Witkowska}},
  \bibinfo{author}{\bibfnamefont{P.}~\bibnamefont{Deuar}},
  \bibinfo{author}{\bibfnamefont{M.}~\bibnamefont{Gajda}}, \bibnamefont{and}
  \bibinfo{author}{\bibfnamefont{K.}~\bibnamefont{Rzazewski}},
  \bibinfo{title}{Solitons as the Early Stage of Quasicondensate Formation during Evaporative Cooling},
  \bibinfo{journal}{Phys. Rev. Lett.} \textbf{\bibinfo{volume}{106}},
  \bibinfo{pages}{135301} (\bibinfo{year}{2011}).

\bibitem[{\citenamefont{Das et~al.}(2012)\citenamefont{Das, Sabbatini, and Zurek}}]{A.Das2012}
  \bibinfo{author}{\bibfnamefont{A.}~\bibnamefont{Das}},
  \bibinfo{author}{\bibfnamefont{J.}~\bibnamefont{Sabbatini}},
  \bibnamefont{and} \bibinfo{author}{\bibfnamefont{W.~H.} \bibnamefont{Zurek}},
  \bibinfo{title}{Winding up superfluid in a torus via Bose Einstein condensation},
  \bibinfo{journal}{Sci. Rep.} \textbf{\bibinfo{volume}{2}},
  \bibinfo{pages}{352} (\bibinfo{year}{2012}).

\bibitem[{\citenamefont{Su et~al.}(2013)\citenamefont{Su, Gou, Bradley, Fialko, and Brand}}]{S.W.Su2013}
  \bibinfo{author}{\bibfnamefont{S.-W.} \bibnamefont{Su}},
  \bibinfo{author}{\bibfnamefont{S.-C.} \bibnamefont{Gou}},
  \bibinfo{author}{\bibfnamefont{A.}~\bibnamefont{Bradley}},
  \bibinfo{author}{\bibfnamefont{O.}~\bibnamefont{Fialko}}, \bibnamefont{and}
  \bibinfo{author}{\bibfnamefont{J.}~\bibnamefont{Brand}},
  \bibinfo{title}{Kibble-Zurek Scaling and its Breakdown for Spontaneous Generation of Josephson Vortices in Bose-Einstein Condensates},
  \bibinfo{journal}{Phys. Rev. Lett.} \textbf{\bibinfo{volume}{110}},
  \bibinfo{pages}{215302} (\bibinfo{year}{2013}).

\bibitem[{\citenamefont{Lamporesi et~al.}(2013)\citenamefont{Lamporesi, Donadello, Serafini, Dalfovo, and Ferrari}}]{G.Lamporesi2013}
  \bibinfo{author}{\bibfnamefont{G.}~\bibnamefont{Lamporesi}},
  \bibinfo{author}{\bibfnamefont{S.}~\bibnamefont{Donadello}},
  \bibinfo{author}{\bibfnamefont{S.}~\bibnamefont{Serafini}},
  \bibinfo{author}{\bibfnamefont{F.}~\bibnamefont{Dalfovo}}, \bibnamefont{and}
  \bibinfo{author}{\bibfnamefont{G.}~\bibnamefont{Ferrari}},
  \bibinfo{title}{Spontaneous creation of Kibble-Zurek solitons in a Bose-Einstein condensate},
  \bibinfo{journal}{Nat. Phys.} \textbf{\bibinfo{volume}{9}},
  \bibinfo{pages}{656} (\bibinfo{year}{2013}).


\bibitem[{\citenamefont{Corman et~al.}(2014)\citenamefont{Corman, Chomaz, Bienaim\'e, Desbuquois, Weitenberg, Nascimb\`ene, Dalibard, and Beugnon}}]{L.Corman2014}
  \bibinfo{author}{\bibfnamefont{L.}~\bibnamefont{Corman}},
  \bibinfo{author}{\bibfnamefont{L.}~\bibnamefont{Chomaz}},
  \bibinfo{author}{\bibfnamefont{T.}~\bibnamefont{Bienaim\'e}},
  \bibinfo{author}{\bibfnamefont{R.}~\bibnamefont{Desbuquois}},
  \bibinfo{author}{\bibfnamefont{C.}~\bibnamefont{Weitenberg}},
  \bibinfo{author}{\bibfnamefont{S.}~\bibnamefont{Nascimb\`ene}},
  \bibinfo{author}{\bibfnamefont{J.}~\bibnamefont{Dalibard}}, \bibnamefont{and}
  \bibinfo{author}{\bibfnamefont{J.}~\bibnamefont{Beugnon}},
  \bibinfo{title}{Quench-Induced Supercurrents in an Annular Bose Gas},
  \bibinfo{journal}{Phys. Rev. Lett.} \textbf{\bibinfo{volume}{113}},
  \bibinfo{pages}{135302} (\bibinfo{year}{2014}).


\bibitem[{\citenamefont{M.Nikkhou et~al.}(2015)\citenamefont{M.Nikkhou et~al.}}]{M.Nikkhou2015}
  \bibinfo{author}{\bibfnamefont{M.}~\bibnamefont{Nikkhou}},
  \bibinfo{author}{\bibfnamefont{M.}~\bibnamefont{Skarabot}},
  \bibinfo{author}{\bibfnamefont{S.}~\bibnamefont{Copar }},
  \bibinfo{author}{\bibfnamefont{M.}~\bibnamefont{Ravnik}},
  \bibinfo{author}{\bibfnamefont{S.}~\bibnamefont{Zumer}},\bibnamefont{and}
  \bibinfo{author}{\bibfnamefont{ I.}~\bibnamefont{Musevic}},
   \bibinfo{title}{Light-controlled topological charge in a nematic liquid crystal},
  \bibinfo{journal}{Nat. Phys.} \textbf{\bibinfo{volume}{11}},
  \bibinfo{pages}{183} (\bibinfo{year}{2015}).


\bibitem[{\citenamefont{J.Sonner et~al.}(2015)\citenamefont{Julian Sonner, Adolfo del Campo, Wojciech H. Zurek}}]{J.Sonner2015}
  \bibinfo{author}{\bibfnamefont{J.Sonner}~\bibnamefont{Nikkhou}},
  \bibinfo{author}{\bibfnamefont{A.}~\bibnamefont{del Campo}}, \bibnamefont{and}
  \bibinfo{author}{\bibfnamefont{W.~H.}~\bibnamefont{Zurek}},
  \bibinfo{title}{Universal far-from-equilibrium dynamics of a holographic superconductor},
  \bibinfo{journal}{Nat. Commun.} \textbf{\bibinfo{volume}{6}},
  \bibinfo{pages}{7406} (\bibinfo{year}{2015}).


\bibitem[{\citenamefont{Zurek et~al.}(2005)\citenamefont{Zurek, Dorner and Zoller}}]{Zurek2005}
  \bibinfo{author}{\bibfnamefont{W.~H.}~\bibnamefont{Zurek}},
  \bibinfo{author}{\bibfnamefont{U.}~\bibnamefont{Dorner}},
  \bibnamefont{and} \bibinfo{author}{\bibfnamefont{P.} \bibnamefont{Zoller}},
  \bibinfo{title}{Dynamics of a Quantum Phase Transition},
  \bibinfo{journal}{Phys. Rev. Lett.}
  \textbf{\bibinfo{volume}{95}}, \bibinfo{pages}{105701}
  (\bibinfo{year}{2005}).


\bibitem[{\citenamefont{Bogdan Damski}(2005)\citenamefont{Bogdan Damski}}]{B.Damski2005}
  \bibinfo{author}{\bibfnamefont{B.}~\bibnamefont{Damski}},
  \bibinfo{title}{The Simplest Quantum Model Supporting the Kibble-Zurek Mechanism of Topological Defect Production: Landau-Zener Transitions from a New Perspective},
  \bibinfo{journal}{Phys. Rev. Lett.}
  \textbf{\bibinfo{volume}{95}}, \bibinfo{pages}{035701}
  (\bibinfo{year}{2005}).

\bibitem{Damski2007}B. Damski and W. H. Zurek, Dynamics of a Quantum Phase Transition in a Ferromagnetic Bose-Einstein Condensate, Phys. Rev. Lett. \textbf{99}, 130402 (2007).

\bibitem[{\citenamefont{Uhlmann et~al.}(2007)\citenamefont{Uhlmann, Sch\"utzhold, and Fischer}}]{M.Uhlmann2007}
  \bibinfo{author}{\bibfnamefont{M.}~\bibnamefont{Uhlmann}},
  \bibinfo{author}{\bibfnamefont{R.}~\bibnamefont{Sch\"utzhold}},
  \bibnamefont{and} \bibinfo{author}{\bibfnamefont{U.~R.}
  \bibnamefont{Fischer}},
  \bibinfo{title}{Vortex Quantum Creation and Winding Number Scaling in a Quenched Spinor Bose Gas},
  \bibinfo{journal}{Phys. Rev. Lett.}
  \textbf{\bibinfo{volume}{99}}, \bibinfo{pages}{120407}
  (\bibinfo{year}{2007}).

\bibitem[{\citenamefont{Dziarmaga et~al.}(2008)\citenamefont{Dziarmaga, Meisner, and Zurek}}]{J.Dziarmaga2008}
  \bibinfo{author}{\bibfnamefont{J.}~\bibnamefont{Dziarmaga}},
  \bibinfo{author}{\bibfnamefont{J.}~\bibnamefont{Meisner}}, \bibnamefont{and}
  \bibinfo{author}{\bibfnamefont{W.~H.} \bibnamefont{Zurek}},
  \bibinfo{title}{Winding Up of the Wave-Function Phase by an Insulator-to-Superfluid Transition in a Ring of Coupled Bose-Einstein Condensates},
  \bibinfo{journal}{Phys. Rev. Lett.} \textbf{\bibinfo{volume}{101}},
  \bibinfo{pages}{115701} (\bibinfo{year}{2008}).

\bibitem[{\citenamefont{Lee}(2009)}]{C.Lee2009}
  \bibinfo{author}{\bibfnamefont{C.}~\bibnamefont{Lee}},
  \bibinfo{title}{Universality and Anomalous Mean-Field Breakdown of Symmetry-Breaking Transitions in a Coupled Two-Component Bose-Einstein Condensate},
  \bibinfo{journal}{Phys. Rev. Lett.} \textbf{\bibinfo{volume}{102}}, \bibinfo{pages}{070401}
  (\bibinfo{year}{2009}).

 \bibitem[{\citenamefont{Sabbatini et~al.}(2011)\citenamefont{Sabbatini, Zurek, and Davis}}]{J.Sabbatini2011}
  \bibinfo{author}{\bibfnamefont{J.}~\bibnamefont{Sabbatini}},
  \bibinfo{author}{\bibfnamefont{W.~H.} \bibnamefont{Zurek}}, \bibnamefont{and}
  \bibinfo{author}{\bibfnamefont{M.~J.} \bibnamefont{Davis}},
  \bibinfo{title}{Phase Separation and Pattern Formation in a Binary Bose-Einstein Condensate},
  \bibinfo{journal}{Phys. Rev. Lett.} \textbf{\bibinfo{volume}{107}},
  \bibinfo{pages}{230402} (\bibinfo{year}{2011}).

\bibitem[{\citenamefont{Chen et~al.}(2011)\citenamefont{Chen, White, Borries, and DeMarco}}]{D.Chen2011}
  \bibinfo{author}{\bibfnamefont{D.}~\bibnamefont{Chen}},
  \bibinfo{author}{\bibfnamefont{M.}~\bibnamefont{White}},
  \bibinfo{author}{\bibfnamefont{C.}~\bibnamefont{Borries}}, \bibnamefont{and}
  \bibinfo{author}{\bibfnamefont{B.}~\bibnamefont{DeMarco}},
  \bibinfo{title}{Quantum Quench of an Atomic Mott Insulator},
  \bibinfo{journal}{Phys. Rev. Lett.} \textbf{\bibinfo{volume}{106}},
  \bibinfo{pages}{235304} (\bibinfo{year}{2011}).

\bibitem[{\citenamefont{Navon et~al.}(2015)\citenamefont{Navon, Gaunt, Smith, and Hadzibabic}}]{N.Navon2015}
  \bibinfo{author}{\bibfnamefont{N.}~\bibnamefont{Navon}},
  \bibinfo{author}{\bibfnamefont{A.~L.} \bibnamefont{Gaunt}},
  \bibinfo{author}{\bibfnamefont{R.~P.} \bibnamefont{Smith}}, \bibnamefont{and}
  \bibinfo{author}{\bibfnamefont{Z.}~\bibnamefont{Hadzibabic}},
  \bibinfo{title}{Critical dynamics of spontaneous symmetry breaking in a homogeneous Bose gas},
  \bibinfo{journal}{Science} \textbf{\bibinfo{volume}{347}},
  \bibinfo{pages}{167} (\bibinfo{year}{2015}).

\bibitem[{\citenamefont{Anquez et~al.}(2016)\citenamefont{Anquez, Robbins, Bharath, Boguslawski, Hoang, and Chapman}}]{M.Anquez2016}
  \bibinfo{author}{\bibfnamefont{M.}~\bibnamefont{Anquez}},
  \bibinfo{author}{\bibfnamefont{B.~A.} \bibnamefont{Robbins}},
  \bibinfo{author}{\bibfnamefont{H.~M.} \bibnamefont{Bharath}},
  \bibinfo{author}{\bibfnamefont{M.}~\bibnamefont{Boguslawski}},
  \bibinfo{author}{\bibfnamefont{T.~M.} \bibnamefont{Hoang}}, \bibnamefont{and}
  \bibinfo{author}{\bibfnamefont{M.~S.} \bibnamefont{Chapman}},
  \bibinfo{title}{Quantum Kibble-Zurek Mechanism in a Spin-1 Bose-Einstein Condensate},
  \bibinfo{journal}{Phys. Rev. Lett.} \textbf{\bibinfo{volume}{116}},
  \bibinfo{pages}{155301} (\bibinfo{year}{2016}).

\bibitem[{\citenamefont{Xu et~al.}(2016)\citenamefont{Jun, Wu, Qin, Huang, Ke, Zhong and Lee}}]{XuJun2016}
  \bibinfo{author}{\bibfnamefont{J.} \bibnamefont{Xu}},
  \bibinfo{author}{\bibfnamefont{S.-Y}~\bibnamefont{Wu}},
  \bibinfo{author}{\bibfnamefont{X.-Z} \bibnamefont{Qin}},
  \bibinfo{author}{\bibfnamefont{J.-H} \bibnamefont{Huang}},
  \bibinfo{author}{\bibfnamefont{Y.-G} \bibnamefont{Ke}},
  \bibinfo{author}{\bibfnamefont{H.-H}
  \bibnamefont{Zhong}},\bibnamefont{and}
  \bibinfo{author}{\bibfnamefont{C.}~\bibnamefont{Lee}},
  \bibinfo{title}{Kibble-Zurek dynamics in an array of coupled binary Bose condensates},
  \bibinfo{journal}{EPL} \textbf{\bibinfo{volume}{113}},
  \bibinfo{pages}{50003} (\bibinfo{year}{2016}).


\bibitem[{\citenamefont{Chin et~al.}(2010)\citenamefont{Chin, Grimm, Julienne, and Tiesinga}}]{C.Chin2010}
  \bibinfo{author}{\bibfnamefont{C.}~\bibnamefont{Chin}},
  \bibinfo{author}{\bibfnamefont{R.}~\bibnamefont{Grimm}},
  \bibinfo{author}{\bibfnamefont{P.}~\bibnamefont{Julienne}}, \bibnamefont{and}
  \bibinfo{author}{\bibfnamefont{E.}~\bibnamefont{Tiesinga}},
  \bibinfo{title}{Feshbach resonances in ultracold gases},
  \bibinfo{journal}{Rev. Mod. Phys.} \textbf{\bibinfo{volume}{82}},
  \bibinfo{pages}{1225} (\bibinfo{year}{2010}).


\bibitem[{\citenamefont{Sachdev}(2010)\citenamefont{Sachdev}}]{Sachdev2011}
  \bibinfo{author}{\bibfnamefont{S.}~\bibnamefont{Sachdev}},
  \bibinfo{title}{Quantum Phase Transition}
  \bibinfo{journal}{(Cambridge Univerisity Press,~Second Edition,~2011)}

\bibitem[{\citenamefont{Robinson}(2010)\citenamefont{Robinson}}]{Robinson2011}
  \bibinfo{author}{\bibfnamefont{M.}~\bibnamefont{Robinson}},
  \bibinfo{title}{Symmetry and the Standard Model}
  \bibinfo{journal}{(Springer-Verlag New York,~2011)}

\bibitem[{\citenamefont{Anatoli Polkovnikov et~al.}(2011)\citenamefont{Anatoli Polkovnikov et al.}}]{A.Polkovnikov2011}
  \bibinfo{author}{\bibfnamefont{A.}~\bibnamefont{Polkovnikov}},
  \bibinfo{author}{\bibfnamefont{K.}~\bibnamefont{Sengupta}},
  \bibinfo{author}{\bibfnamefont{A.}~\bibnamefont{Sengupta}}, \bibnamefont{and}
  \bibinfo{author}{\bibfnamefont{M.}~\bibnamefont{Vengalattore}},
  \bibinfo{title}{\emph{Colloquium}: Nonequilibrium dynamics of closed interacting quantum systems},
  \bibinfo{journal}{Rev. Mod. Phys.} \textbf{\bibinfo{volume}{83}},
  \bibinfo{pages}{863} (\bibinfo{year}{2011}).


\bibitem[{\citenamefont{Giorgini et~al.}(2008)\citenamefont{Giorgini, Pitaevskii, and Stringari}}]{S.Giorgini2008}
  \bibinfo{author}{\bibfnamefont{S.}~\bibnamefont{Giorgini}},
  \bibinfo{author}{\bibfnamefont{L.~P.} \bibnamefont{Pitaevskii}},
  \bibnamefont{and}
  \bibinfo{author}{\bibfnamefont{S.}~\bibnamefont{Stringari}},
  \bibinfo{title}{Theory of ultracold atomic Fermi gases},
  \bibinfo{journal}{Rev. Mod. Phys.} \textbf{\bibinfo{volume}{80}},
  \bibinfo{pages}{1215} (\bibinfo{year}{2008}).

\bibitem[{\citenamefont{Ha et~al.}(2013)\citenamefont{Ha, Hung, Zhang, Eismann, Tung, and Chin}}]{L.C.Ha2013}
  \bibinfo{author}{\bibfnamefont{L.-C.} \bibnamefont{Ha}},
  \bibinfo{author}{\bibfnamefont{C.-L.} \bibnamefont{Hung}},
  \bibinfo{author}{\bibfnamefont{X.}~\bibnamefont{Zhang}},
  \bibinfo{author}{\bibfnamefont{U.}~\bibnamefont{Eismann}},
  \bibinfo{author}{\bibfnamefont{S.-K.} \bibnamefont{Tung}}, \bibnamefont{and}
  \bibinfo{author}{\bibfnamefont{C.}~\bibnamefont{Chin}},
  \bibinfo{title}{Strongly Interacting Two-Dimensional Bose Gases},
  \bibinfo{journal}{Phys. Rev. Lett.} \textbf{\bibinfo{volume}{110}},
  \bibinfo{pages}{145302} (\bibinfo{year}{2013}).

\bibitem[{\citenamefont{Bakr et~al.}(2010)\citenamefont{Bakr, Peng, Tai, Ma, Simon, Gillen, F{\"o}lling, Pollet, and Greiner}}]{W.S.Bakr2010}
  \bibinfo{author}{\bibfnamefont{W.~S.} \bibnamefont{Bakr}},
  \bibinfo{author}{\bibfnamefont{A.}~\bibnamefont{Peng}},
  \bibinfo{author}{\bibfnamefont{M.~E.} \bibnamefont{Tai}},
  \bibinfo{author}{\bibfnamefont{R.}~\bibnamefont{Ma}},
  \bibinfo{author}{\bibfnamefont{J.}~\bibnamefont{Simon}},
  \bibinfo{author}{\bibfnamefont{J.~I.} \bibnamefont{Gillen}},
  \bibinfo{author}{\bibfnamefont{S.}~\bibnamefont{F{\"o}lling}},
  \bibinfo{author}{\bibfnamefont{L.}~\bibnamefont{Pollet}}, \bibnamefont{and}
  \bibinfo{author}{\bibfnamefont{M.}~\bibnamefont{Greiner}},
  \bibinfo{title}{Probing the Superfluid{\textendash}to{\textendash}Mott Insulator Transition at the Single-Atom Level},
  \bibinfo{journal}{Science} \textbf{\bibinfo{volume}{329}},
  \bibinfo{pages}{547} (\bibinfo{year}{2010}).

\bibitem[{\citenamefont{Sherson et~al.}(2010)\citenamefont{Sherson, Weitenberg, Endres, Cheneau, Bloch, and Kuhr}}]{J.F.Sherson2010}
  \bibinfo{author}{\bibfnamefont{J.~F.} \bibnamefont{Sherson}},
  \bibinfo{author}{\bibfnamefont{C.}~\bibnamefont{Weitenberg}},
  \bibinfo{author}{\bibfnamefont{M.}~\bibnamefont{Endres}},
  \bibinfo{author}{\bibfnamefont{M.}~\bibnamefont{Cheneau}},
  \bibinfo{author}{\bibfnamefont{I.}~\bibnamefont{Bloch}}, \bibnamefont{and}
  \bibinfo{author}{\bibfnamefont{S.}~\bibnamefont{Kuhr}},
   \bibinfo{title}{Single-atom-resolved fluorescence imaging of an atomic Mott insulator},
  \bibinfo{journal}{Nature} \textbf{\bibinfo{volume}{467}}, \bibinfo{pages}{68}
  (\bibinfo{year}{2010}).

\end{thebibliography}
\end{document}